\documentclass{JHEP3}
\usepackage[utf8]{inputenc}
\usepackage{amstext}
\usepackage{graphicx}
\usepackage{amssymb}
\usepackage{esint}
\usepackage{longtable}

\newcommand{\be}{\begin{equation}}
\newcommand{\ee}{\end{equation}}

\newcommand{\mincir}{\raise
-3.truept\hbox{\rlap{\hbox{$\sim$}}\raise4.truept\hbox{$<$}\ }}
\newcommand{\magcir}{\raise
-3.truept\hbox{\rlap{\hbox{$\sim$}}\raise4.truept\hbox{$>$}\ }}
\newcommand{\ba}{\begin{eqnarray}}
\newcommand{\ea}{\end{eqnarray}}
\newcommand{\brr}{\begin{array}}
\newcommand{\err}{\end{array}}
\newcommand{\bc}{\begin{center}}
\newcommand{\ec}{\end{center}}

\title{Precision growth index using the clustering of cosmic structures and growth data}

\author{%
Athina Pouri$^{a,b}$, Spyros Basilakos$^{a}$, Manolis Plionis$^{c,d,e}$\\

$^{a}$ Academy of Athens, Research Center for Astronomy and Applied
Mathematics, Soranou Efesiou 4, 11527, Athens, Greece\\

$^{b}$ Faculty of Physics, Department of Astrophysics - Astronomy -
Mechanics, University of Athens, Panepistemiopolis, Athens 157 83 \\

$^{c}$ Physics Dept., Sector of Astrophysics, Astronomy \& Mechanics, Aristotle Univ. of Thessaloniki,
Thessaloniki 54124, Greece \\

$^{d}$ Instituto Nacional de Astrof\'isica \'Optica y Electronica, 72000 Puebla, M\'exico \\

$^{e}$ IAASARS, National Observatory of Athens, P.Pendeli 15236, Greece \\

E-Mails: \email{athpouri@phys.uoa.gr}, \email{svasil@academyofathens.gr},
\email{mplionis@auth}}

\abstract{
We use the clustering properties of Luminous Red Galaxies (LRGs)
and the growth rate data provided by the various galaxy surveys
in order to constrain the growth index ($\gamma$) of the linear
matter fluctuations. We perform a standard $\chi^2$-minimization procedure between theoretical expectations and data, followed by a joint likelihood
analysis and we find
a value of $\gamma=0.56\pm 0.05$, perfectly consistent
with the expectations of the $\Lambda$CDM model, and
$\Omega_{m0} =0.29\pm 0.01$, in very good agreement
with the latest Planck results. Our analysis
provides significantly more stringent growth index constraints
with respect to previous studies,
as indicated by the fact that the corresponding uncertainty
is only $\sim 0.09 \gamma$. Finally, allowing $\gamma$ to vary with
redshift in two manners (Taylor expansion around $z=0$, and Taylor
expansion around the scale factor), we find that the combined
statistical analysis between our
clustering and literature growth data
alleviates the degeneracy
and obtain more stringent constraints with respect to other recent studies.
}

\begin{document}

\section{Introduction}
The statistical analysis of various cosmological data
(SNIa, Cosmic Microwave Background-CMB, Baryonic Acoustic
Oscillations-BAOs, Hubble parameter measurements etc) strongly
suggests that we live in a spatially flat universe
that consists of $\sim 4\%$ baryonic matter, $\sim 26\%$
dark matter and $\sim 70\%$ some sort of dark energy (hereafter DE) which is
necessary to explain the accelerated expansion of the
universe (cf. \cite{Hicken2009,Komatsu2011,Blake2011,Hinshaw2013,Farooq2013,Ade2013,Sperg2013} and references therein).
Although there is a common agreement regarding the
ingredients of the universe, there are different views
concerning the possible physical mechanism which is responsible
for the cosmic acceleration. Briefly, the general path that one can follow
in order to mathematically treat the accelerated expansion of the
universe is to see DE either as a new field in nature or as a
modification of General Relativity (see for review
\cite{Copeland2006,Caldwell2009,Amendola2010}).

An interesting approach to discriminate between scalar field
DE and modified gravity is to use the evolution of the
linear growth of matter perturbations
$\delta_{m} (z)=\delta \rho_m / \rho_m$
\cite{Linder2004,Linder2007,Mar2014}.
Specifically, a useful tool in this kind of studies is the
so called growth rate of clustering,
which is defined as
$f(a)=\frac{d{\rm ln}D}{d{\rm ln}a}\simeq \Omega_{m}^{\gamma}(a)$,
where $a(z)=(1+z)^{-1}$ is the scale factor of the universe,
$\Omega_m(a)$ is the dimensionless matter density parameter, $\gamma$
is the growth index and $D(a)=\delta_{m}(a)/\delta_{m}(a=1)$
is the linear growth factor scaled to unity at the present time
\cite{Peebles1993,Wang1998}.
The accurate determination of the growth index is considered one of
the main goals of Observational
Cosmology because it can be used in order to check the validity
of General Relativity (GR) on cosmological scales.
The basic ingredient in this approach comes from the fact that
$\gamma$ depends weakly on the dark energy equation of state (hereafter EoS)
parameter w$(z)$ \cite{Linder2007},
implying that one can split the background
expansion history, $H(z)$, constrained by
geometric probes (SNIa, BAO, CMB), from the dynamical
perturbations growth history. 

Theoretically speaking, it has been shown that for those DE
models which are within the framework of GR and have
a constant EoS parameter, the growth index $\gamma$ is
well approximated by $\gamma \simeq \frac{3({\rm w}-1)}{6{\rm w}-5}$
\cite{Linder2007,Silveira1994,Wang1998,Nesseris2008}.
In the case of the concordance
$\Lambda$CDM model (w$(z)=-1$) the above formula reduces to
$\gamma \approx 6/11$. Considering
the braneworld model of \cite{DGP2000} we have $\gamma \approx 11/16$
(see \cite{Linder2007,Wei2008,Gong2008,Fu2009}).
Finally,
for some $f(R)$ gravity models it has been found that
$\gamma \simeq 0.415-0.21z$
for various parameter values (see
\cite{Gannouji2009,Tsujikawa2009}),
while for the Finsler-Randers
cosmology, Basilakos \& Stavrinos \cite{Basilakos2013}
found $\gamma \approx 9/14$.

From the large scale structure point of view, the study of
the distribution of matter on
large scales using different extragalactic mass tracers
(galaxies, AGNs, clusters
of galaxies etc) provides important constraints on
structure formation theories.
In particular, since gravity reflects, via gravitational instability,
on the nature of clustering \cite{Peebles1993}
it has been proposed to use
the clustering/biasing properties
of the mass tracers in constraining cosmological models
(see \cite{Matsubara2004,Basilakos2005,Basilakos2006,
Krumpe2013})
as well as to test
the validity of GR on extragalactic scales
(\cite{Basilakosetal2012} for a recent review see \cite{Bean2013}).

Based on the above arguments, the aim of the current study is
to place constraints on the
$(\Omega_{m}, \gamma)$ parameter space using
the measured two-point angular correlation function (hereafter ACF)
of the LRGs, 
with known redshift
distribution. The basic idea is to compare the measured and the theoretically
predicted ACF, which is based on the 3D power spectrum, the evolution of the relevant
bias parameter and the Limber's inversion integral equation.
Therefore, the predicted theoretical ACF depends also on the growth of
linear matter perturbations via which the $(\Omega_{m},\gamma)$ pair,
which can thus be constrained.
The merit of utilizing ACF data for such a task is related to
the fact that we do not need to consider a fiducial cosmological
model in order to derive the ACF data, as well as on the fact that the ACF is
unaffected by redshift-space distortions.
In addition to ACF, we use the recent growth rate data
as collected by Nesseris \& Garcia-Bellido \cite{NesGa12}, Hudson
\& Turnbull \cite{Hud12}, Beutler et al. \cite{Beutler} and Basilakos et al.
\cite{BasilakosNes2013} in order to put tight constraints
on $(\Omega_{m},\gamma)$.


The structure of the paper is as follows.
In section 2 we present the angular correlation function data, measured for LRGs and the growth data.
In section 3 we discuss the theoretical
angular correlation function model and basic ingredients in order to
calculate it, such as
the linear growth of matter perturbations, the evolution of the linear
bias factor and the CDM power spectrum.
The details of our methodology used to fit models to the data and our
results are presented in section 4, while our main conclusions in section
5.

\section{Angular Correlation Function Data and Growth data}
It is well known that the two-point ACF, $w(\theta)$, is defined as
the excess joint probability over random
of finding two mass tracers (galaxies, AGNs, clusters)
separated by an angular separation
$\theta$. Therefore by definition we have $w(\theta)=0$
for a random distribution of sources.
In this work we use the ACF of 2SLAQ LRGs galaxies
with median redshift $z_{\star}\simeq 0.55$.
In particular, we utilize the ACF of 655775 photometrically selected LRGs
from the SDSS DR5 catalogue, already estimated in
\cite{Sawangwit2011}.
This sample has been
compiled using the same selection criteria as the
2dF-SDSS LRG and Quasar survey (hereafter 2SLAQ), which covers the
redshift range: $0.45<z<0.8$.
Following the original paper of \cite{Sawangwit2011} we use the ACF
up to an angular scale of $6000^{''}$ in order to avoid the effects of
BAO's. Since the aim of our paper is to put constraints on the
\underline{linear} growth index we also exclude
small angular scales ($\theta<140^{''}$, which corresponds to $\le 1
\; h^{-1}$ Mpc at $z_{\star}$) where strong non-linear
effects (the so-called ``one-halo'' term) are expected, although we do
use in our theoretical modeling a mildly non-linear correction term
(see also section 3.3).
In Table 1 we list the precise numerical
values of the ACF data points with the corresponding errors that are
used in our analysis.

In addition, we utilize in our analysis the growth rate of clustering
data which are based on
the PSCz, 2dF, VVDS, SDSS, 6dF, 2MASS, BOSS and {\em WiggleZ} galaxy surveys,
for which their combination parameter of the growth rate of structure,
$f(z)$, and the redshift-dependent rms fluctuations of the linear
density field, $\sigma_8(z)$,
is available as a function of redshift, $f(z)\sigma_{8}(z)$.
The total sample contains 16 entries
(as collected by Basilakos et al. \cite{BasilakosNes2013}).
The $f\sigma_{8}$ estimator is almost a model-independent
way of expressing the observed growth history
of the universe (see \cite{Song09}).
Indeed the observed growth rate
of structure ($f_{obs}=\beta b$) is derived
from the redshift space distortion parameter $\beta(z)$ and the linear
bias.
Observationally, using the anisotropy of the spatial
correlation function one can estimate the $\beta(z)$
parameter (see also section 3.2).
On the other hand, the linear bias factor can be defined as
the ratio of the variances of the tracer (galaxies, QSOs etc) and underlying mass density fields, smoothed at $8h^{-1}$ Mpc
$b(z)=\sigma_{8,tr}(z)/\sigma_{8}(z)$,
where $\sigma_{8,tr}(z)$ is measured directly
from the sample. Combining the above definitions
we arrive at $f \sigma_{8}=\beta \sigma_{8,tr}$. 
Since different authors have estimated $f \sigma_{8}$
using different cosmologies, we need to convert
them to the same cosmological background in order to
be able to utilize them consistently.
Specifically, we wish to translate the value of growth data $f\sigma_{8}$
from a reference cosmological model,
say Ref, to the background cosmology.
The definition of $f(z)\simeq \Omega_{m}(z)^{\gamma(z)}$
and $\sigma_{8}(z)=\sigma_{8}D(\Omega_{m0},z)$ (for more details see section 3.1) simply implies a correction factor:
\be
\label{CCF}
C_{f}=\frac{f\sigma_{8,\rm obs}}{f\sigma_{8,\rm obs}^{\rm Ref}}=
\left[\frac{\Omega_{m}(z)}{\Omega_{m}^{\rm Ref}(z)}\right]^{\gamma(z)}
\frac{\sigma_{8}D(\Omega_{m0},z)}{\sigma_{8}^{\rm
    Ref}D(\Omega_{m0}^{\rm Ref},z)} \;.
\ee
Notice that the $f\sigma_{8,\rm obs}^{\rm Ref}$ data
and the corresponding uncertainties 
can be found in Table 1 
of \cite{BasilakosNes2013}.


\begin{table}
\caption[]{The measured angular correlation function data
of the 2SLAQ LRGs from \cite{Sawangwit2011}. We use here
bootstrap errors meaning that we need to
multiply the uncertainties of \cite{Sawangwit2011} with $\sqrt{3}$.}
\tabcolsep 10pt
\begin{tabular}{cccc}
\hline
Index & $\theta^{''}$ & $w(\theta)$ & $\delta w(\theta)$\\ \hline \hline
1& 153.72 & 0.285 & 0.0061\\
2& 230.64 & 0.199 & 0.0038\\
3& 345.96 &  0.152 & 0.0026\\
4& 518.94 &  0.113 & 0.0019\\
5& 778.2 &  0.078 & 0.0018\\
6& 1167.6 & 0.055 &  0.0012\\
7& 1751.4 &  0.038 & 0.0011\\
8& 2626.8 &  0.0226 & 0.0009\\
9& 3600 &  0.0144 & 0.0008\\
10& 4800 &  0.0086 & 0.00076\\
11& 6000 & 0.0054  & 0.00067\\
\end{tabular}
\end{table}


\section{Modeling the Theoretical Correlation Function}
In this section we briefly discuss the basic steps of modeling the
theoretically expected ACF for the two different mass tracers used
and of the evolution of bias of extragalactic mass
tracers. Considering a spatially flat
Friedmann-Lema\^\i tre-Robertson-Walker (FLRW) geometry), we can
easily relate via the Limber's inversion equation the ACF with the two
point spatial correlation function $\xi(r,z)$:
\be
\label{wwa}
w(\theta)=2\frac{H_0}{c} \int\limits_0^\infty
\left(\frac{1}{N}\frac{dN}{dz}\right)^2 E(z)dz\int\limits_0^\infty
{\xi (r,z)du} \;,
\ee
where $1/N \; dN/dz$ is the normalized redshift distribution of the
sources under study, 
which is provided by a random subsample of the source population for
which redshifts (spectroscopic or photometric) are available
(see section 4).

The spatial correlation function
of the mass tracers is given by
\begin{equation}
\label{ssep2}
\xi(r,z)=b^2(z)\xi_{DM}(r,z)
\end{equation}
where $b(z)$ is the evolution of the linear bias,
and $\xi_{DM}$ is the corresponding correlation function of the
underlying mass distribution which is written as
\begin{equation}
\label{ssep1}
\xi_{DM}(r,z)=\frac{1}{2\pi^2}\int \limits_0^\infty k^{2} P(k,z)
\frac{{\rm sin}(kr/a)}{(kr/a)}dk    \;.
\end{equation}
with $P(k,z)=D^2(z) P(k)$, and $P(k)$ denoting the power spectrum
of the matter fluctuations.

The variable $r$ corresponds to the
physical separation between two sources
having an angular separation, $\theta$ (in steradians). In the case of a
small angle approximation the physical separation becomes
\be
\label{ssep}
r \simeq a(z)\left(u^{2}+x^{2}\theta^{2} \right)^{1/2}
\ee
where $u$ is the line-of-sight separation of any two sources and
$x(z)$ is the comoving distance, given by:
\be
x(z)=\frac{c}{H_{0}}\int_{0}^{z} \frac{dy}{E(y)} \;,
\ee
$E(z)=H(z)/H_0$, is the normalized Hubble parameter.

Inserting Eqs.(\ref{ssep2}), (\ref{ssep1}), (\ref{ssep}) and $a(z)=1/(1+z)$
into Eq.(\ref{wwa}) and integrating over the variable $u$
we arrive at (see also \cite{Cress1998}) our final
theoretically expected ACF:
\be
\label{wwep}
w(\theta)=\frac{1}{2\pi}
\int_{0}^{\infty} k^{2}P(k)dk \int_{0}^{\infty}
D^2(z)
j(k,z,\theta) dz
\ee
with
\be
j(k,z,\theta)=\frac{H_{0}}{c}\left(\frac{1}{N}\frac{dN}{dz}\right)^2
b^2(z)E(z)J_{0}(k\theta x(z))
\ee
where $J_0$ is the Bessel function of zero kind, given by:
\be J_{0}(\omega)=\frac{2}{\pi}\int_{0}^{\infty} {\rm
  sin}\left(\omega\;{\rm cosh}\tau\right)d\tau
=\frac{1}{\pi}\int_{0}^{\pi} {\rm cos}\left(\omega\;{\rm
    sin}\tau\right)d\tau
\ee
Note that the expectations for the different mass tracers enter
through the bias evolution factor, $b(z)$, and the tracer redshift
distribution $1/N \;dN/dz$.

Obviously, the dependence of ACF on gravity as well as on
the different expansion models enters through the
behavior of $D(z)$, which in turn depends on
$\gamma$ (see equation \ref{Dz222}),
and on $E(z)=H(z)/H_{0}$ respectively.
In the subsections below we present the different ingredients that
enter in Eq.(\ref{wwep}), namely, the linear perturbation growth rate,
the bias evolution factor and the CDM power spectrum.

\subsection{The linear growth factor $D(a)$ and the linear growth rate $f(a)$}
Here we provide the form of the linear density perturbation growth
rate, which is the ingredient through which the growth index,
$\gamma$, enters in our analysis.

At sub-horizon scales the basic
differential equation which describes the linear matter fluctuations
(\cite{Linder2004,Linder2007,Lue2004,Stabenau2006,Uzan2007,Tsujikawa2008} 
and references therein) is
\be
\label{odedelta}
\ddot{\delta}_{m}+ 2H\dot{\delta}_{m}=4 \pi G_{\rm eff} \rho_{m} \delta_{m}
\ee
where $\rho_{m}\propto a^{-3}$ is the matter density,
$G_{\rm eff}= G_N Q(t)$ with $G_N$ being the
Newton's gravitational constant and the function $Q(t)$ depends on
gravity.
For the scalar field DE models,
$G_{\rm eff}$ is equal to $G_{N}$, i.e., $Q(a)=1$, while for
the case of modified gravity models
we have $Q(a) \ne 1$ and subsequently $G_{\rm eff}\ne G_{N}$.

The growing-mode solution of equation (\ref{odedelta}) is
$\delta_{m}\propto D(a)$, where $D(a)$ is the linear growing mode
usually scaled to unity at the present time.
Generally, for either modified gravity or scalar field DE
we can write the following useful parametrization \cite{Peebles1993,
Wang1998,Linder2007}
\be \label{fa}
 f(a)=\frac{d{\rm ln} \delta_m}{d{\rm ln}a} \simeq \Omega_{m}^{\gamma}(a)
\ee
where $\Omega_{m}(a)=\Omega_{m0}a^{-3}/E^{2}(a)$.
Therefore, using $d/dt=H \; d/d\ln a$ we express
Eq.(\ref{odedelta}) in terms of $f(a)$ as:
\be
\label{faaa}
\frac{df}{d{\rm ln}a}+ f^2 +\left(\frac{\dot H}{H^2}+2\right)f=
\frac{3}{2} Q(a)  \Omega_{m}(a)
\ee
where for the $\Lambda$CDM expansion we have
\be
\frac{\dot H}{H^{2}}+2=\frac{1}{2}-\frac{3}{2}{\rm
  w}(a)\left[1-\Omega_{m}(a)\right]
\ee
and w$(a)=-Q(a)=-1$. In this case the
normalized Hubble parameter $E(a)$, is:
\be
E(a)=(\Omega_{m0}a^{-3}+\Omega_{\Lambda 0})^{1/2}
\ee
with $\Omega_{\Lambda 0}=1-\Omega_{m0}$ and $H_{0}$ the Hubble
constant\footnote{For the comoving distance
  and for the dark matter halo mass we use the traditional
  parametrization $H_{0}=100h$km/s/Mpc. Of course, when we treat the
  power spectrum shape parameter $\Gamma$ we utilize
  $h\equiv {\tilde h}=0.68$ \cite{Sperg2013}.}.
As we have stated in the introduction we can separate the
background expansion $H(a)$ from the growth history \cite{Linder2007}.

The parametrization of Eq.(\ref{fa}) greatly
simplifies the numerical calculations
of Eq.(\ref{odedelta}). Indeed, providing a direct integration of
Eq.(\ref{fa}) we easily find
\begin{equation}
\label{Dz221}
\delta_{m}(a,\gamma)=a(z) \;{\rm exp} \left[\int_{a_{i}}^{a(z)} \frac{dy}{y}\;
\left(\Omega_{m}^{\gamma}(y)-1\right) \right]
\end{equation}
where $a_{i}$ is the scale factor of the universe
at which the matter component dominates the cosmic fluid
(here we use $a_{i} \simeq 10^{-2}$). Then the linear
growth factor, normalized to unity at the present epoch,
is:
\begin{equation}
\label{Dz222}
D(a)=\frac{\delta_{m}(a,\gamma)}{\delta_{m}(1,\gamma)}=\frac{a(z)
  \;{\rm exp} \left[\int_{a_{i}}^{a(z)} \frac{dy}{y}\;
\left(\Omega_{m}^{\gamma}(y)-1\right) \right]}
{{\rm exp} \left[\int_{a_{i}}^{1} \frac{dy}{y}\;
\left(\Omega_{m}^{\gamma}(y)-1\right) \right]} \;.
\end{equation}

However, $\gamma$ may not be a constant but rather evolve with
redshift; $\gamma\equiv \gamma(z)$. In such a case, inserting
Eq.(\ref{fa}) into Eq.(\ref{faaa}) we obtain:
\begin{equation}
\label{agamz}
-(1+z)\gamma^{\prime}{\rm ln}(\Omega_{m})+\Omega_{m}^{\gamma}+ 3{\rm w}   \label{Poll}
(1-\Omega_{m})\left(\gamma-\frac{1}{2}\right)+\frac{1}{2}=\frac{3}{2}Q\Omega_{m}^{1-\gamma}
\end{equation}
where the prime denotes derivative with respect to redshift.
Various functional forms of $\gamma(z)$ have been proposed in the
literature \cite{Polarski2008,Bal08,Belloso2009,Basilakos2012},
for example:
\begin{equation}
\gamma(z)=\left\{ \begin{array}{cc}
       \gamma_{0}+\gamma_{1}z, &
       \mbox{$\Gamma_{1}$-parametrization}\\
       \gamma_{0}+\gamma_{1}z/(1+z),& \mbox{$\Gamma_{2}$-parametrization.}
       \end{array}
        \right.
\end{equation}
Using the above parametrizations and Eq.(\ref{agamz}) evaluated at the
present time ($z=0$), one can easily obtain
the parameter $\gamma_{1}$ in terms of $\gamma_{0}$
\begin{equation}
\label{Poll2}
\gamma_{1}=\frac{\Omega_{m0}^{\gamma_{0}}+3{\rm w}_{0}(\gamma_{0}-\frac{1}{2})
(1-\Omega_{m0})-\frac{3}{2}Q_{0}\Omega_{m0}^{1-\gamma_{0}}+\frac{1}{2}  }
{\ln  \Omega_{m0}}\;.
\end{equation}
Owing to the fact that the $\Gamma_{1}$
parametrization is valid only at relatively low redshifts ($0\le z \le
0.5$), for $z>0.5$ we utilize $\gamma=\gamma_{0}+0.5\gamma_{1}$.
As an example, in the case of the usual $\Lambda$CDM cosmological model
(ie., $Q_{0}=1$, ${\rm w}_{0}=-1$ and $\gamma_{0}^{(th)}\simeq 6/11$) with
$\Omega_{m0}=0.30$, Eq.(\ref{Poll2}) gives
$\gamma_{1}^{(th)}\simeq -0.0459$.

\subsection{The evolution of linear bias, $b(z)$}
Here we briefly present the model that we use to trace the evolution of the
linear bias factor, which reflects the relation between the
overdensities of luminous and of dark matter
\cite{Kaiser1984,Bardeen1986}.
We remind the reader that biasing is considered to be statistical
in nature with galaxies and clusters being identified as high peaks
of an underlying, initially Gaussian, random density field.
The usual paradigm is of a linear and scale-independent bias,
defined as the ratio of density perturbations in the mass-tracer field
to those of the underline total matter field: $b = \delta_{tr}/\delta_m$
\footnote{
We would like to point that up
to galaxy cluster scales the fluctuations of the metric do not
introduce a significant scale dependence in the growth factor
\cite{Dent2009} and in the linear bias \cite{Basilakosetal2012}.}.

In this analysis we use the bias evolution model
of \cite{Basilakosetal2012,Basilakos2011}.
This generalized model is  based on the linear perturbation theory and
the Friedmann-Lemaitre solutions of the cosmological field equations.
It is valid for any DE model (scalar or geometrical) and it is given by:
\be
b(z)=1+\frac{b_{0}-1}{D(z)}+C_{2} \frac{{\cal J}(z)}{D(z)}
\ee
with
\be
{\cal J}(z) = \int_{0}^{z} {\frac{(1 + y)}{E(y)}dy} \;.
\ee
The constants $b_0$ (the bias at the present time)
and $C_{2}$ depend on the host dark matter halo mass, as we have
verified using $\Lambda$CDM N-body simulations (see \cite{Basilakosetal2012}), and are given by:
\be
b_0(M_{h})=0.857 \left[ 1+
  \left(C_m \;\frac{M_{h}}{10^{14}\;h^{-1}M_{\odot}}\right)^{0.55} \right]
\ee
\be
C_2(M_{h})=1.105 \left(C_m \;\frac{M_{h}}{10^{14}\;h^{-1}M_{\odot}}\right)^{0.255} \;,
\ee
where $C_m=\Omega_{m0}/0.27$.



\subsection{CDM Power Spectrum, $P(k)$}
The CDM power spectrum is given by $P(k)=P_{0} k^{n}T^{2}(k)$,
where $T(k)$ is the CDM transfer function
and $n\simeq 0.9671$ following the recent reanalysis of the Planck
data by Spergel et al. \cite{Sperg2013}. Regarding $T(k)$,
we use two different functional forms namely, that of
Bardeen et al. \cite{Bardeen1986} and of Eisenstein \& Hu \cite{Eisenstein1998}.

The \cite{Bardeen1986} is given by:
\be
\label{jtf}
T(k)=C_{q}\Big[1+3.89 q + (16.1 q)^2 + (5.46 q)^3+(6.71 q)^4\Big]^{-1/4}
\ee
where $C_{q}=\frac{{\rm ln}(1+2.34 q)}{2.34q}$ and
$q\equiv \frac{k}{\Gamma}$.
Here $\Gamma$ is the shape parameter, given according to
\cite{Sugiyama1995} as:
\be
\Gamma=\Omega_{m0}{\tilde h}{\rm
  exp}(-\Omega_{b0}-\sqrt{2\tilde{h}}\;\Omega_{b0}/\Omega_{m0})\;.
\ee
The value of $\Gamma$, which is kept constant throughout the model
fitting procedure, is estimated using
the Planck results of Spergel el al. \cite{Sperg2013}\footnote{We use
the Planck priors provided by Spergel et al. \cite{Sperg2013} in order
to avoid possible systematics on the cosmological parameters
which are related to the problematic (according to Spergel et al.)
217GHz$\times$217GHz detector. However, at the end of the analysis we
provide results based on the Planck results of Ade et al. \cite{Ade2013}.}
namely,
$\Omega_{b0}=0.022197{\tilde h}^{-2}$, ${\tilde h}=0.68$ and
$\Omega_{m0}=0.302$.
The alternative transfer function used is that of \cite{Eisenstein1998}:
\be
T(k)=\frac{L_{0}}{L_{0}+C_{0}q^{2}}
\ee
where $L_{0}={\rm ln}(2e+1.8q)$, $e=2.718$ and $C_{0}=14.2+\frac{731}{1+62.5q}$.

Also, the rms fluctuations of the linear density
field on mass scale $M_{h}$ is:
\be
\label{ssig}
\sigma(M_{h},z)=\left[\frac{D^{2}(z)}{2 \pi^{2}}\int_{0}^{\infty} k^{2}P(k)
W^{2}(kR) dk \right]^{1/2} \;,
\ee
where $W(kR)=3({\rm sin}kR-kR{\rm cos}kR)/(kR)^{3}$ and
$R=(3M_{h}/ 4\pi \rho_{0})^{1/3}$ with
$\rho_{0}$ denotes the mean matter density of the
universe at the present time
($\rho_{0}=2.78 \times 10^{11}\Omega_{m0}h^{2}M_{\odot}$Mpc$^{-3}$).
To this end, the normalization of the power spectrum is given by:
\be
P_{0}=2\pi^{2} \sigma_{8}^{2} \left[ \int_{0}^{\infty} T^{2}(k)
 k^{n+2} W^{2}(kR_{8})dk \right]^{-1}
\ee
where $\sigma_{8}\equiv \sigma(R_{8},0)$ is the rms mass fluctuation
on $R_{8}=8 h^{-1}$ Mpc scales and for which we use separately 
either of the two following parametrizations:
\begin{equation}\label{ss88}
\sigma_{8}=\left\{ \begin{array}{cc}
       0.818\left( \frac{0.30}{\Omega_{m0}}\right)^{0.26}, &
       \mbox{Spergel et al. \cite{Sperg2013}}\\
 0.797\left( \frac{0.30}{\Omega_{m0}}\right)^{0.26},&
\mbox{Hajian et al. \cite{Hajian2013}} .
       \end{array}
        \right.
\end{equation}
Finally, we would like to stress that
we have taken into account the non-linear corrections by using
the corresponding fitting formula
introduced by \cite{Peacock1994}, for the $\Lambda$CDM model 
(see also \cite{Smith2003,Widrow2009}). 
In their fitting formula there is one relatively free
parameter, which is the slope of the
power spectrum at the relevant scales, because the CDM power
spectrum curves slowly and thus it varies as a function of
scale according to: $n_{\rm eff}=d{\rm ln}P/d{\rm ln}k$.

An alternative model for the non-linear power spectrum is provided by the halofit model, presented in Smith et al. \cite{Smith2003} 
(see also \cite{Tak2012}), as an advancement
to that of Peacock \& Dodds \cite{Peacock1994}, with its main novelty 
being its decomposition into two terms, the halo-halo term 
and the one-halo term (modeling better the smaller scale power-spectrum), 
and its applicability to more general spectra. 
However, as it is evident from the analysis of 
Smith et al. (\cite{Smith2003}: see their Fig.14), 
the convolution of the two terms provide an overall
non-linear power spectrum which is consistent with 
that of Peacock \& Dodds \cite{Peacock1994}
mostly for the CDM type spectra. 
This is especially true for the range of interest in our work, ie., 
$k<1 h$ Mpc$^{-1}$ (which corresponds to angular 
separations $\theta>140^{''}$ at $z_{\star}=0.55$), 
where the two models provide similar non-linear power spectrums.



\begin{table*}
\caption[]{Results in the $(\Omega_{m0},\gamma,M_{h},n_{\rm eff})$ parameter space for the
  different $T(k)$ and $\sigma_{8}$.}
\tabcolsep 12pt
 \begin{tabular}{cccccc} \hline
$T(k)$ & $\Omega_{m0}$ & $\gamma$ & $M_h/10^{13} M_{\odot}$ &$n_{\rm eff}$ & $\chi^2_{t, {\rm min}}/df$ \\ \hline \\
\multicolumn{6}{c}{$\sigma_{8}=0.797\left( 0.30/\Omega_{m0}\right)^{0.26}$ \cite{Hajian2013}}\\
& &  & & & \\
Eisenstein \& Hu \cite{Eisenstein1998} & $0.29 \pm 0.01$ & $0.56\pm 0.05$ &
$1.90\pm 0.10$ & $0.10\pm 0.20$ & 16.36/23\\
Bardeen et al. \cite{Bardeen1986}    &$0.29\pm 0.01$ & $0.56\pm 0.10$ &
$1.80\pm 0.30$ & $ -0.10^{+0.30}_{-0.10}$ & 16.56/23 \\
\hline \\
\multicolumn{6}{c}{$\sigma_{8}=0.818\left(0.30/\Omega_{m0}\right)^{0.26}$ \cite{Sperg2013}}\\
& &  & & &\\
Eisenstein \& Hu \cite{Eisenstein1998} & $0.29^{+0.03}_{-0.02}$ &$0.58^{+0.02}_{-0.06}$ & $1.70\pm 0.20$ & $0.30\pm 0.20$ & 15.90/23\\
 Bardeen et al. \cite{Bardeen1986}     & $0.29^{+0.02}_{-0.03}$ & $0.56 \pm 0.10$ &$1.60\pm 0.4$ & $0.0_{-0.20}^{+0.10}$ & 16.13/23 \\

\end{tabular}
\end{table*}

\begin{figure}
\begin{center}
\includegraphics[width=0.8\textwidth]{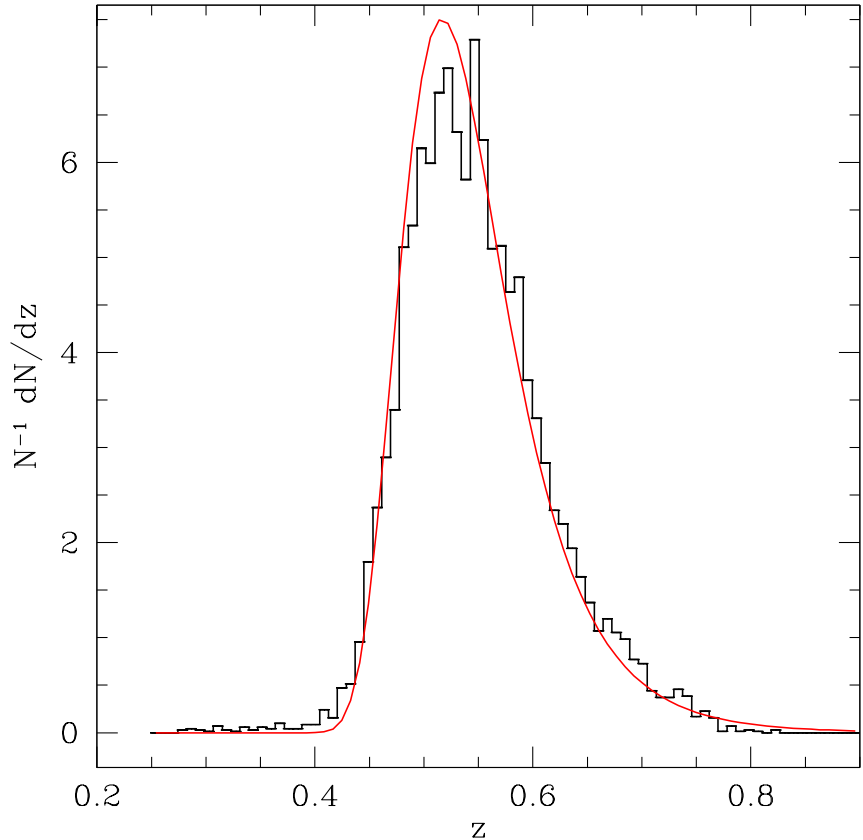}
\caption{The normalized photometric redshift distribution
of the 2SLAQ LRG galaxies.
The red continuous
line is its corresponding best fit according to Eq.(4.1).}
\end{center}
\end{figure}

\begin{figure}
\begin{center}
\includegraphics[width=0.8\textwidth]{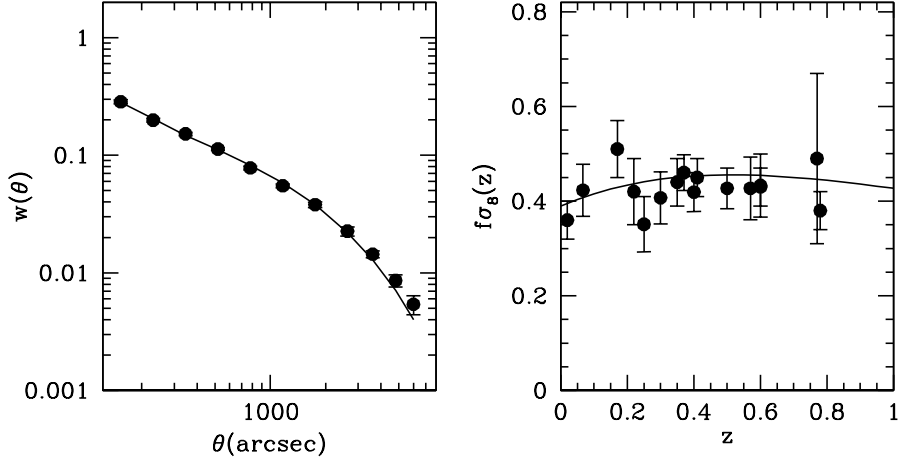}
\caption{{\it Left Panel:} Comparison of the observed (solid points)
and theoretical angular correlation function. 
{\it Right Panel:} Comparison of
the observed (solid points) and theoretical evolution of the growth
rate $f(z)\sigma_{8}(z)$. In order to obtain the theoretical curve we use
$(\Omega_{m0},\gamma)=(0.29,0.56)$ (for more details see section 4.1).
}
\end{center}
\end{figure}

\begin{figure}
\begin{center}
\includegraphics[width=0.8\textwidth]{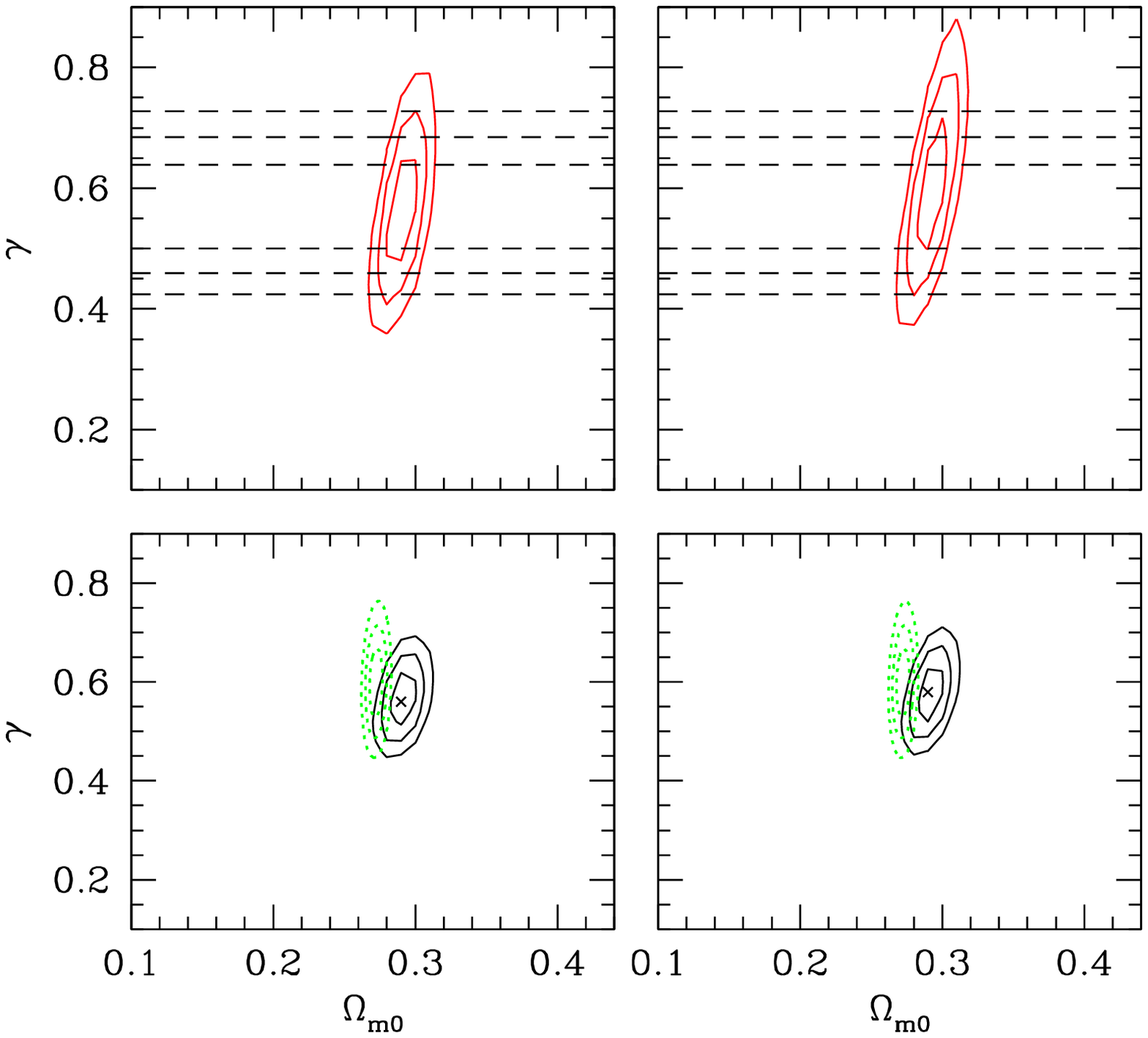}
\caption{Likelihood contours for
$\Delta \chi^2=\chi^{2}_{t}-\chi^{2}_{t,{\rm min}}$
equal to 2.32, 6.18 and 11.83, corresponding to
1$\sigma$, 2$\sigma$ and $3\sigma$ confidence levels, in the
$(\Omega_{m0},\gamma)$ plane using the Eisenstein \& Hu \cite{Eisenstein1998}
transfer function
and the \cite{Basilakosetal2012,Basilakos2011} bias model.
{\it Left Panels:} The likelihood contours correspond to
Hajian et al. \cite{Hajian2013} power
spectrum normalization. {\it Right Panels:}
The contours here correspond to $\sigma_{8}$ provided by
Spergel et al. \cite{Sperg2013}.
The best fit solutions 
are represented by the crosses. Note that using the Bardeen et al.
\cite{Bardeen1986} transfer function we find almost the same
results within $1\sigma$ errors.
In the upper panels we present the likelihood
contours that correspond to the LRGs (solid red lines) and growth data
(dashed black lines). Notice, that in order to plot 
the LRG contours  
we have marginalized over $M_{h}$ and $n_{\rm eff}$ (see Table 2). 
The bottom 
panels show the joint statistical results.
Finally, the
green dotted curves are the SNIa/BAOs/${\rm CMB}_{\rm shift}$/$f\sigma_{8}$
joint likelihood contours provided by \cite{Nesseris2013}.
}
\end{center}
\end{figure}

\section{Fitting Theoretical Models to the data}
In this section we implement a standard $\chi^{2}$ minimization
statistical analysis in order to provide constraints either in the
 $(\Omega_{m0},\gamma)$ parameter space, or for $\gamma(z)$.
An important ingredient that is necessary in
Eq.(\ref{wwa}), in order to relate the spatial to the angular two-point 
correlation functions, is the tracer redshift distribution.
For the LRGs
we use an analytic model of their photometric
redshift distribution, which we then insert in Eq.(\ref{wwa}).
The model redshift distribution is given by fitting the
following useful formula to the data:
\be
\label{NNcl}
\frac{dN}{dz}\propto \left( {\frac{z}{{z_{\star}}}} \right)^{(a+2)}
e^{ - \left( {\frac{z}{{z_{\star}}}} \right)^{\beta} } \;.
\ee
We obtain the relevant parameters by fitting the data of the redshift distribution to the above formula:

\be
(a, \beta, z_{\star})=(-15.53, -8.03, 0.55)
\ee
where $z_{\star}$ is the characteristic depth of the subsample studied.
In Fig.1, we present the estimated normalized redshift distribution
$(\frac{1}{N}\frac{dN}{dz})$ and the corresponding continuous fit provided by
Eq.(\ref{NNcl}).

We are now set to compare the measured 2SLAQ LRGs and growth
functions with the predictions of different spatially flat $\Lambda$
cosmological models.
To this end we use the standard $\chi^2$-minimization procedure, which in
our case it is defined as follows:

\noindent
{\bf (1)} For the LRG clustering cosmological probe:
\be
\label{eq:likel}
\chi^{2}_{\rm LRGs} ({\bf p}_{1},{\bf p}_{2})=
\sum_{i=1}^{11} \frac{\left[ w_{\rm th} (\theta_{i},{\bf
      p}_{1},{\bf p}_{2})-w_{obs}(\theta_{i}) \right]^{2}}
{\sigma^{2}_{i}}  \;\;.
\ee
where the expected theoretical ACF ($w_{\rm th}$) is given by Eq.(\ref{wwep}) and $\sigma_i$ is the observed ACF 1$\sigma$ uncertainty, and

\noindent
{\bf (2)} for the growth-rate cosmological probe:
\be
\label{Likel}
\chi^{2}_{\rm gr}({\bf p}_{1})=
\sum_{i=1}^{16} \left[ \frac{C_{f}(z_{i},{\bf p}_{1})f\sigma_{8,\rm
      obs}^{\rm Ref}(z_{i})-
f\sigma_{8}(z_{i},{\bf p}_{1})}
{C_{f}(z_{i},{\bf p}_{1}) \sigma_{i}^{\rm Ref}}\right]^{2}
\ee
where $\sigma_{i}^{\rm Ref}$ is the observed 1$\sigma$ uncertainty, while
$C_{f}$ is given in Eq.(\ref{CCF}), and the theoretical growth-rate is given by:
\be
f\sigma_{8}(z,{\bf p}_{1})=\sigma_{8}D(z)\Omega_{m}(z)^{\gamma(z)}\;.
\ee
The vectors ${\bf p}_{1}$ and ${\bf p}_{2}$ provide the
free parameters that enter in deriving the theoretical expectations.
The ''cosmo-gravity'' ${\bf p}_{1}$ vector contains those
free parameters which are related to
the expansion and gravity. For the case of constant $\gamma$ it is
defined as: ${\bf p}_{1}=(\Omega_{m0},\gamma,\sigma_{8})$, and for the
case of evolving $\gamma$, as: ${\bf p}_{1}=(\Omega_{m0},\gamma_0,
\gamma_1, \sigma_{8})$.
The ${\bf p}_{2}=(M_{h},n_{\rm eff})$ vector is
associated with the environment of the dark matter halo in which
the extragalactic mass tracers (in our case LRGs galaxies) live.



Since we wish to perform a joint likelihood analysis of the two cosmological
probes and since likelihoods are defined
as ${\cal L}\propto {\rm exp}(-\chi^{2}/2)$, one has that the
joint likelihood is:
\begin{equation}\label{eq:overalllikelihoo1}
{\cal L}_{t} ({\bf p}_{1},{\bf p}_{2})=
{\cal L}_{\rm LRGs}({\bf p}_{1},{\bf p}_{2})
\times {\cal L}_{\rm gr}({\bf p}_{1}) \;,
\end{equation}
which is equivalent to:
\begin{equation}\label{eq:overalllikelihoo}
\chi^{2}_{t}({\bf p}_{1},{\bf p}_{2})=
\chi^{2}_{\rm LRGs}({\bf p}_{1},{\bf p}_{2})+
\chi^{2}_{\rm gr}({\bf p}_{1}) \;.
\end{equation}
Based on the above we will provide our results for each free parameter that
enters in the two ${\bf p}_{1,2}$ vectors. Note that the uncertainty of each
fitted parameter will be estimated after marginalizing one
parameter over the other, providing as its uncertainty the
range for which $\Delta \chi^{2}(\le 1\sigma)$.
Such a definition, however,
may hide the extent of a possible degeneracy between the
fitted parameters and thus it is important to visualize the solution
space, as indicated in the relevant contour figures.

As a further consistency check we have used the inverse of
the Fisher matrix, the covariance matrix, but we find similar
uncertainties to those provided by the marginalization method, most probably
due to the fact that the 1, 2 and 3$\sigma$ solution space contours are
symmetric and the axes of symmetry are parallel to the ${\bf p}_{1,2}$ vectors.
Since the errors of the Fisher matrix approach
 are symmetric by definition, we have decided to use the marginalization
approach.

In the left panel of Fig. 2, we present the observed $w(\theta)$ for the 2SLAQ
LRGs (left panel),
with the best fit model of the angular correlation function provided
by Eq.(\ref{wwep}) and the minimization procedure discussed above.
In the right panel of Fig. 2,
we plot the growth data (solid points) as collected by Basilakos et al.
(see \cite{BasilakosNes2013}
and references therein) with the estimated
(solid line)
growth rate function, $f(z)\sigma_{8}(z)$ (for more details see the discussion
section 4.1).


\begin{table*}
\caption[]{Literature growth results for the $\Lambda$CDM cosmological model.
The last line corresponds to our results. Similar
to \cite{BasilakosNes2013} 
results can be also found in \cite{Nesseris2013}.} 
\tabcolsep 4.5pt
 \begin{tabular}{cccc}

\hline
\hline

Data used                     & $\Omega_{m0}$       & $\gamma$                   & References\\ \hline \hline

galaxy data from 2dFGRS   & $0.30 \pm 0.02$  & $0.60_{-0.30}^{+0.40}$        & \cite{diPorto2008}\\ 

old $f(z)$ growth data        & 0.30              & $0.674_{-0.169}^{+0.195}$  & \cite{Nesseris2008}\\ 

old $f(z)$ growth data        & $0.273\pm 0.015$ & $0.64_{-0.15}^{+0.17}$     & \cite{Gong2008}\\ 

X-ray cluster luminocity function+$f_{gas}$            &  $0.214^{+0.036}_{-0.041}$   & $0.42^{+0.20}_{-0.16}$   & \cite{Rapetti2010}\\ 

WMAP+SNIa+MCMC            &  0.25            & $0.584 \pm 0.112 $       & \cite{Samushia2012}\\  

old+new $f(z)$ growth data & $0.273\pm 0.011$   & $0.586_{-0.074}^{+0.079} $ & \cite{Basilakos2012}\\   

$f \sigma_8 $ growth data & $0.259\pm 0.045$ & $0.619 \pm 0.054 $         & \cite{Hudson2012}\\  

$f \sigma_8 $ growth data & 0.273            & $0.602 \pm 0.055$          & \cite{BasilakosP2012}\\    

old+new $f(z)$ growth data & 0.273  & $0.58 \pm 0.04$ & \cite{Lee2012}\\    

$f \sigma_8 $ growth data+(SNIa, BAOs, ${\rm CMB}_{\rm shift}$)  &  $0.272 \pm 0.003$ & $0.597 \pm 0.046 $ &
\cite{Nesseris2013,BasilakosNes2013}\\  

CMASS DR9+ other $f\sigma_8$  & $ 0.308 \pm 0.022 $   & $0.64 \pm 0.05 $ & \cite{Samushia2013}\\    


cl+CMB+gal+SNIa+BAO & $0.284 \pm 0.012$  & $0.618\pm 0.062$ & \cite{Rapetti2013}\\   

Lensing + $f\sigma_{8}$ growth data & $0.256 \pm 0.023$  & $0.52\pm 0.09$ & \cite{Simpson2013}\\   

CMB+clustering of Baryon Oscillation Spec. Survey& $0.30 \pm 0.01$  & $0.69\pm 0.15$ & \cite{Sann2013}\\   

$f \sigma_8 $ growth data+(SNIa, BAOs, ${\rm CMB}_{\rm CAMB}$)  &  $0.298^{+0.027}_{-0.023}$ & $0.675^{+0.18}_{-0.16}$ &
\cite{XULIN}\\  

{\bf Clustering of LRGs+ growth data}  &  $0.29\pm 0.01$ & $0.56\pm 0.05$ & {\bf Our study}\\

\hline
\end{tabular}
\end{table*}

\subsection{Constraints on $(\Omega_{m0}, \gamma)$}
In our analysis we have set $\sigma_{8}$ based on Eq.(\ref{ss88})
and thus the cosmogravity vector contains only two independent
free parameters, namely ${\bf p}_{1}=(\Omega_{m0},\gamma)$. Thus, we have 
four free parameters in total.
We sample the various parameters as follows:
the matter density $\Omega_{m0} \in [0.1,1]$ in steps of
0.01; the growth index $\gamma \in [0.1,1.0]$ in steps
of 0.01, the parent dark matter halo (for the LRGs)
$M_h/10^{13}h^{-1}M_{\odot}$ $\in [1,2.5]$
and the slope of the
power spectrum $n_{\rm eff}\in [-0.5,1.0]$ in steps of 0.1.


In Table 2 we present our resulting parameter joint constraints separately
for the case of the
Hajian et al. \cite{Hajian2013} and the
Spergel et al. \cite{Sperg2013}
power spectrum normalizations respectively (see Eq.(\ref{ss88})),
as well as for the two different CDM transfer functions used.

A first general result is that the two transfer functions used provide
very similar cosmo-gravity results within $1\sigma$ errors.
Therefore, in the rest of the paper we
utilize the Eisenstein \& Hu \cite{Eisenstein1998} transfer function.
Secondly, we would like to mention that
the $\chi^2_{t,{\rm min}}$ for the Hajian et al. \cite{Hajian2013}
normalization,
results
in a reduced value of $\chi^{2}_{t,{\rm min}}/df\sim 16.36/23$ while
the corresponding $\chi^{2}_{t,{\rm min}}/df$ value for
the Spergel et al. \cite{Sperg2013} is $\sim 15.90/23$.
In Figure 3 we present the 1$\sigma$, 2$\sigma$ and $3\sigma$
confidence contours in the $(\Omega_{m0},\gamma)$ plane for
both $\sigma_{8}$ normalization
(\cite{Sperg2013}: right panel and \cite{Hajian2013}: left panel).
These results are based
on the transfer function of Ref.\cite{Eisenstein1998}.  

In the upper panels of Fig.3 we present 
the likelihood contours for the 
individual sets of data on LRGs (solid red contours) and 
growth data (dashed black contours), whereas in the bottom panels of 
Fig. 3 we display the corresponding combined likelihood contours.
One can see from Fig.3 (upper panels) that the growth 
data place constraints on $\gamma$, however the value of $\Omega_{m}$ is not 
constrained by the growth analysis and
all the values in the interval $0.1 \le \Omega_{m} \le 1$ are acceptable
within the $1\sigma$ uncertainty. In contrast, the value of $\Omega_{m}$ 
is well defined using the statistical analysis of LRGs.

As it can also be seen from Table 2, 
$\Omega_{m0}=0.29\pm 0.01$, which is in a very 
good agreement with the 
Planck \cite{Sperg2013} results, while the 
derived value of $\gamma=0.56\pm 0.05$
coincides with the theoretically expected $\Lambda$CDM value.
The tight joint constraints come from the fact that 
the individual contours (upper panels of Fig3) are vertical.
Inserting $\Omega_{m0}=0.29$ into the second branch of Eq.(\ref{ss88}) we obtain
$\sigma_{8} \simeq 0.804$. The aforementioned environmental vector is
${\bf p}_{2}=\left( (1.90\pm 0.2) \times 10^{13}h^{-1}M_{\odot}, 0.10\pm 0.20\right)$.
It is interesting to mention, that our derived
host DM halo mass is consistent with that of
Sawangwit et al. \cite{Sawangwit2011}, namely
$M_{h}=(2.1\pm 0.1) \times 10^{13}h^{-1}M_{\odot}$.
Alternatively, considering the Planck prior \cite{Sperg2013}
of $\Omega_{m0}=0.30$
and minimizing with respect to $\gamma$ and
${\bf p}_{2}=(M_{h},n_{\rm eff})$
we find $\gamma=0.56\pm 0.05$ and
${\bf p}_{2}=\left( (2.0\pm 0.10) \times 10^{13}h^{-1}M_{\odot}, 0.30\pm 0.20\right)$ with $\chi^{2}_{t,{\rm min}}/df\sim 16.52/24$.



With respect to other recent studies, our best fit values of $\gamma$
are in agreement, within $1\sigma$ errors, to that of \cite{Nesseris2013}
(see also \cite{BasilakosNes2013}) who found
$\gamma=0.597 \pm 0.046 $, using a combined statistical analysis
of expansion and growth data (SNIa/BAOs/${\rm CMB}_{\rm shift}$/$f\sigma_{8}$).
However, our joint $\Omega_{m0}$ value
is somewhat greater
(within $\sim 1.8\sigma$ uncertainty), from the derived value of
\cite{Nesseris2013}, $\Omega_{m0}=0.272 \pm 0.003$\footnote{Nesseris et al.
\cite{Nesseris2013} imposed $\sigma_{8}=0.80$.}.
For comparison, in Figure 3 we also display the combined $(\Omega_{m0},\gamma)$
likelihood contours (see green dotted lines) of \cite{Nesseris2013}.
It is evident that the combined analysis of the growth data with
the LRGs clustering provides strong constraints on the growth
parameter $\gamma$, which implies that this method
works equally well with that of the joint
SNIa/BAOs/${\rm CMB}_{\rm shift}$/$f\sigma_{8}$.

In order to appreciate the great effort in the recent years to
estimate jointly $\Omega_{m0}$ and $\gamma$ and the relative strength
and precision of the different methods, we present a summary
of relevant literature results in Table 3. 
It appears unavoidable to conclude that current
data favor, within a $1\sigma$ uncertainty, the theoretically predicted value of
$\gamma_{\Lambda}^{(th)}\simeq 6/11$.
Secondly, the quality and quantity
of the cosmological and dynamical (growth and the like)
data as well as methodologies have greatly improved in recent years.
For example,
since the first measurement of \cite{diPorto2008}, the  
error budget of the growth index has been decreased by
one order of magnitude
with respect to best fit value of the current work.
It is also important to note that
using only the combine basic properties
of the large scale structures (ACF of 2SLAQ LRGs and growth data) we have
managed to significantly reduce the growth index uncertainty,
namely $\sigma_{\gamma}/\gamma\sim 9\%$ and thus producing
one of the strongest (to our knowledge) existing joint constraints on $\gamma$.

\begin{figure}
\begin{center}
\includegraphics[width=0.8\textwidth]{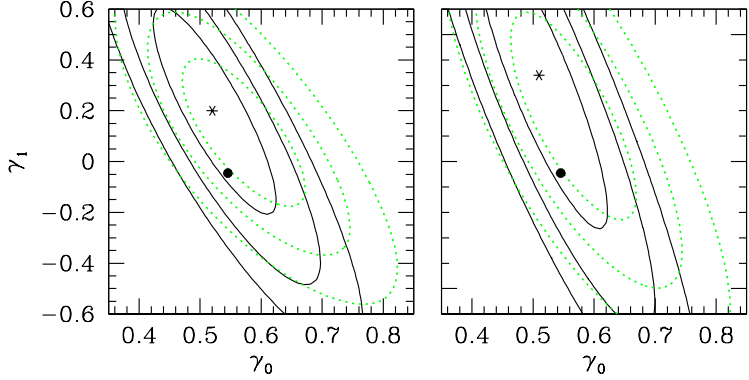}
\caption{The joint 2SLAQ (LRGs)
galaxy $w(\theta)$ and $f\sigma_{8}$
  likelihood contours (solid curves)
in the ($\gamma_{0},\gamma_{1}$) plane
(using $\Omega_{m0}=0.30$ and $\sigma_{8}=0.797$).
The left and right panels show the
results based on the 
$\Gamma_{1}$ and $\Gamma_2$ parametrizations respectively.
The crosses correspond to the best fit parameters. We also
show the theoretical $\Lambda$CDM
$(\gamma_{0}^{(th)},\gamma_{1}^{(th)})$ values (solid points)
given in section 3.1.
Finally, the
green dotted curves are the SNIa/BAOs/${\rm CMB}_{\rm shift}$/$f\sigma_{8}$
joint likelihood contours provided by \cite{Nesseris2013}.
}
\end{center}
\end{figure}

\begin{table*}
\caption[]{Literature $(\gamma_{0},\gamma_{1})$ constraints. The bold phase
lines correspond to the present analysis.}
\tabcolsep 6pt
\begin{tabular}{cccc}
\hline
Parametrization Model & $\gamma_0$ & $\gamma_1$ & Reference\\ \hline \hline

$\Gamma_{1}:\;\;\gamma(z)=\gamma_{0}+\gamma_{1}z$ & 0.77 $\pm$ 0.29 & -0.38 $\pm$ 0.85 & \cite{Nesseris2008}\\ 

 & $0.774$ & $-0.556$  & \cite{Fu2009}\\    

 & $0.49_{-0.11}^{+0.12}$ & $0.305_{-0.318}^{+0.345}$  & \cite{Basilakos2012}\\   

 & 0.48 $\pm$ 0.07 & 0.32 $\pm$ 0.20 & \cite{Lee2012}\\   

 & $0.40_{-0.080}^{+0.086}$ & $0.603\pm 0.241$  & \cite{BasilakosP2012}\\  

 & $0.567 \pm 0.066$ & $0.116\pm 0.19$  & \cite{Nesseris2013,BasilakosNes2013}\\  

 & $0.520 \pm 0.04$ & $0.02\pm 0.11$  & {\bf Our study} \\ 
 &  & &  \\
$\Gamma_{2}:\;\;\gamma(z)=\gamma_{0}+\gamma_{1}z/(1+z)$& $0.92_{-1.26}^{+1.56}$ & $-1.49_{-6.08}^{+6.86}$  & \cite{Dosset2010}\\  

 & $0.461_{-0.11}^{+0.12}$ & $0.513_{-0.414}^{+0.448}$  & \cite{Basilakos2012}\\   

 & 0.46 $\pm$ 0.09 & 0.55 $\pm$ 0.36 & \cite{Lee2012}\\    

 & $0.345_{-0.080}^{+0.085}$ & $1.006\pm 0.314$  & \cite{BasilakosP2012}\\   

 & $0.561 \pm 0.068$ & $0.183\pm 0.26$  & \cite{Nesseris2013,BasilakosNes2013}\\ 


 & $0.560 \pm 0.03$ & $-0.10\pm 0.11$  & {\bf Our study} \\ 

\hline

\end{tabular}
\end{table*}

\subsection{Constraints on $\gamma(z)$}

In this section we perform a consistent minimization procedure in the
$(\gamma_0,\gamma_1)$ parameter space.
Following the considerations discussed in the previous section
and for the sake of simplicity the ''cosmo-gravity'' vector becomes
${\bf p}_{1}=(0.30,\gamma_{0},\gamma_{1},\sigma_{8})$, where
we have marginalized the overall likelihood analysis over
the LRG environmental vector
$(M_{h},n_{\rm eff})=\left( 2.0 \times 10^{13}h^{-1}M_{\odot}, 0.30 \right)$.
We sample $\gamma_{0} \in [0.35,0.85]$ in steps of 0.01 and
$\gamma_{1} \in [-0.6,0.6]$ in steps of 0.01.
Note, that as in \cite{Nesseris2013} we first
use a constant $\sigma_{8}$, namely $\sigma_{8}=0.797$.

\begin{figure}
\begin{center}
\includegraphics[width=0.8\textwidth]{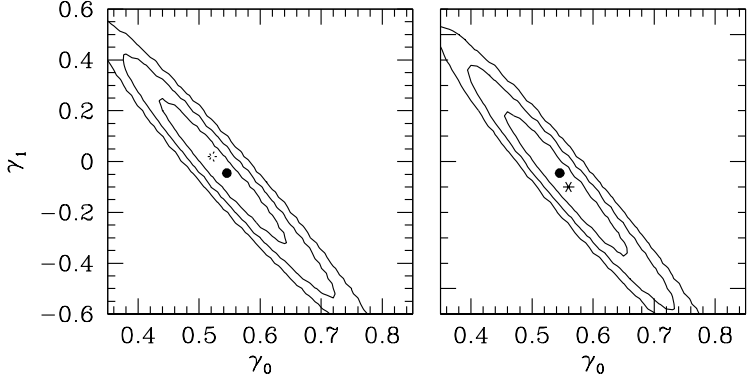}
\caption{The likelihood contours in the $(\gamma_{0},\gamma_{1})$
plane (see caption of Fig. 4 for definitions).
Here the normalization of the power spectrum is given by
Eq.(4.8).
}
\end{center}
\end{figure}

In Fig.4 we plot the results of our statistical analysis in the
$(\gamma_{0},\gamma_{1})$ plane for the
Eisenstein \& Hu \cite{Eisenstein1998} transfer function, since we
have verified that using
Bardeen et al. \cite{Bardeen1986} transfer function
we get similar contours. The left panel shows the results based on the
$\Gamma_{1}$ parametrization while
the right panel those of the $\Gamma_{2}$ parametrization.
Our contours are in agreement with those of
\cite{Nesseris2013} (see green dashed curves in Fig.4)
which implies that practically
our likelihood analysis provides similar results with those of
SNIa/BAOs/${\rm CMB}_{\rm shift}$/$f\sigma_{8}$.

The theoretical $(\gamma_{0}^{(th)},\gamma_{1}^{(th)})$
$\Lambda$CDM values (see above) are indicated by the solid points while
the stars represent our best fit values which are:
\begin{itemize}
  \item for the $\Gamma_1$ parametrization we have
$\chi^{2}_{t,\rm min}/df=16.0/25$,
$\gamma_0=0.52 \pm 0.08$, $\gamma_1=0.20 \pm 0.32$.
\item for the $\Gamma_2$ parametrization:
similarly, we obtain
$\chi^{2}_{t,\rm min}/df=15.87/25$,
$\gamma_0=0.51 \pm 0.08$, $\gamma_1=0.34^{+0.26}_{-0.46}$.
\end{itemize}

Obviously, the $\gamma_{1}$ parameter is not constrained by
this analysis, which is however also the case for the joint
SNIa/BAOs/${\rm CMB}_{\rm shift}$/$f\sigma_{8}$ analysis \cite{Nesseris2013}.
This effect is partially attributed to the constant $\sigma_{8}$.
Therefore, we attempt to alleviate this by
additionally treating the $\sigma_{8}$ prior properly along the
$\gamma$-chain. Following the normalization procedure
of \cite{BLM10} we rescale the value of $\sigma_{8}$ by
\be
\label{sLL}
\sigma_{8,\gamma}=\sigma_{8} \frac{\delta_{m}(1,\gamma_{0},\gamma_{1})}{
\delta_{m}(1,\gamma_{0}^{(th)},\gamma_{1}^{(th)})} \;.
\ee
where $\sigma_{8}=0.797$ and $\delta_{m}(a,\gamma)$ is given by Eq.(\ref{Dz221}).
We repeat our statistical analysis by using $\sigma_{8,\gamma}$
in the ''cosmo-gravity'' vector and we find:
\begin{itemize}
  \item for the $\Gamma_1$ parametrization:
$\gamma_0=0.52 \pm 0.04$, $\gamma_1=0.02 \pm 0.11$ with
$\chi^{2}_{t,\rm min}/df=17.5/25$.
\item for the $\Gamma_2$ parametrization:
$\gamma_0=0.56 \pm 0.03$, $\gamma_1=-0.10 \pm 0.11$
with $\chi^{2}_{t,\rm min}/df \simeq 17.3/25$.
\end{itemize}

It is evident
that the predicted $\Lambda$CDM
$(\gamma_{0}^{(th)},\gamma_{1}^{(th)})$ values
of both parametrizations
are close to the best fit parameters
(see solid points in Fig. 5).
Finally, comparing
the contours of Fig.5 with literature results
(for the corresponding Refs. see Table 3)
we find that indeed we have managed to reduce significantly the
area of $\gamma_{0}-\gamma_{1}$ contours, increasing the Figure of
merit by 
$\sim 30\%$, with respect to that of the constant $\sigma_{8}$
analysis (see Fig. 4).
Also, in Table 4, one may see a more compact presentation
of our results including literature best fit $(\gamma_{0},\gamma_{1})$
values.


\subsection{Using the priors provided by the Planck team}
In order to complete the current study
we repeat our analysis by using those
priors derived originally by the Planck team \cite{Ade2013}, namely
$$(\Omega_{b0},{\tilde h},n,\sigma_{8})=
(0.02207{\tilde h}^{-2},0.674,0.9616,\sigma_{8})$$
with $\sigma_{8}=0.87(0.27/\Omega_{m0})^{0.3}$.
Since, we have found that the results remain mostly unaffected by using
the two different forms of $T(k)$, we use here
the form of Ref.\cite{Eisenstein1998}.
In brief we find:

\begin{itemize}

\item the overall likelihood function peaks at
$(\Omega_{m0},\gamma)=(0.29^{+0.02}_{-0.03},0.56^{+0.02}_{-0.06})$
with $\chi^2_{t,\rm min}/df\simeq 15/23$.
The corresponding environmental vector is
${\bf p}_{2}=
(1.40 \pm {0.1}\times 10^{13}\;h^{-1}M_{\odot},0.4\pm 0.20)$.
If we impose $\Omega_{m0}=0.315$ then we find $\gamma=0.58^{+0.02}_{-0.10}$, 
${\bf p}_{2}=
(1.60 \pm {0.1}\times 10^{13}\;h^{-1}M_{\odot},0.8\pm 0.10)$ with 
$\chi^2_{t,\rm min}/df\simeq 17.1/24$. 

Furthermore, using the latter $\Omega_{m0}$ and ${\bf p}_{2}$ we obtain:
\item in the case of $\Gamma_1$ parametrization:
$\chi^{2}_{t,\rm min}/df=18/25$,
$\gamma_0=0.55 \pm 0.04$, $\gamma_1=-0.06 \pm 0.12$.

\item in the case of $\Gamma_2$ parametrization:
$\chi^{2}_{t,\rm min}/df=17.8/25$,
$\gamma_0=0.55 \pm 0.04$, $\gamma_1=-0.08 \pm 0.12$.
Notice, that for both $\gamma(z)$ parametrizations we utilize Eq.(\ref{sLL})
as far as the variable $\sigma_{8,\gamma}$ is concerned.

\end{itemize}

\section{Conclusions}
In the epoch of intense cosmological studies aimed
at testing the validity of general relativity on extragalactic scales, it
is very important to minimize the amount of priors
needed to successfully complete such an effort.
One such prior is the growth index
and its measurement at the $\sim 1\%$
accuracy level has been proposed as a necessary step
for checking possible departures from general relativity
at cosmological scales \cite{Bean2013}.
Therefore, it is of central importance
to have independent determinations of $\gamma$, because
this will help to control the systematic
effects that possibly affect individual methods and tracers of the
growth of matter perturbations.

In this study we use the basic large scale structure properties
such as the clustering
of the 2SLAQ Luminous Red Galaxies together with
the growth rate of clustering provided by the various galaxy surveys
in order to constrain the growth index of matter
perturbations. The results of the two analyzes are used
in a joint likelihood fitting procedure
which helps to reduce the parameter uncertainties.
The outcome constraints are: $(\Omega_{m0},\gamma)=(0.29\pm 0.01,0.56\pm 0.05)$,
which are the strongest (to our knowledge)
joint constraints appearing in the literature.
Also, we check that our growth results are quite robust
against the choice of the transfer function of the power spectrum
and the Planck priors which are available in the literature
\cite{Ade2013,Sperg2013}.

Finally, considering a time varying growth index:
$\gamma(z)=\gamma_{0}+\gamma_{1}X(z)$,
with $X(z)=z$ or
$X(z)=z/(1+z)$
we find, as all similar studies, that
$\gamma_{1}$ and $\gamma_{0}$ are somehow degenerate.
However, based on the joint statistical
analysis we have managed
to put tighter constraints on $\gamma_{0}$.
Although, we have reduced significantly the $\gamma_{1}$
uncertainty with respect to previous studies, the corresponding error bars
remain quite large.
Future, dynamical data
are expected to improve even further the relevant constraints
(especially on $\gamma_{1}$) and thus the validity of
GR on cosmological scales will be effectively tested.

\vspace{0.0cm}

\acknowledgments We are greatly thankful to S. Nesseris for
providing us with an electronic
version of their growth $\Omega_{m0}-\gamma$ and $\gamma_0 - \gamma_1$ contours.
We also thank D. G. Ballesteros and C. Marinoni for useful 
comments and suggestions.
AP acknowledges financial support under
the Academy of Athens: {\it Fellowship for Astrophysics} grant 2005-49878.
SB also acknowledges support by the Research Center for
Astronomy of the Academy of Athens
in the context of the program  ``{\it Tracing the Cosmic Acceleration}''.



\end{document}